\documentclass[prl,twocolumn,amsmath,amssymb,floatfix,tighten,bibtex]{revtex4-1}
\usepackage{graphicx}
\usepackage{bm}
\usepackage{amssymb}
\usepackage{color}
\usepackage{epsfig}
\usepackage{subfigure}

\newcommand{\be}{\begin{equation}}
\newcommand{\ee}{\end{equation}}
\begin{document}

\title {Helical liquids and Majorana bound states in quantum wires}

\author{Yuval Oreg,$^1$ Gil Refael,$^2$ and Felix von Oppen$^3$}

\affiliation{$^1$Department of Condensed Matter Physics, Weizmann
Institute of Science, Rehovot, 76100, Israel }
\affiliation{$^2$Department of Physics, California Institute of Technology, Pasadena, California 91125, USA}
\affiliation{$^3$Dahlem Center for Complex Quantum Systems and Fachbereich Physik, Freie Universit\"at Berlin, 14195 Berlin, Germany}
\begin{abstract}
We show that the combination of spin-orbit coupling with a Zeeman field or strong interactions may lead to the formation of a helical liquid in single-channel quantum wires. In a helical liquid, electrons with opposite velocities
have opposite spin precessions. We argue that zero-energy Majorana bound states are formed in various situations
when the wire is situated in proximity to a conventional $s$-wave superconductor. This occurs when the external magnetic field, the superconducting gap, or the chemical potential vary along the wire. We discuss experimental consequences of the formation of the helical liquid and the Majorana bound states.
\end{abstract}

 \maketitle

States of matter that support Majorana fermions have received much attention in the context of quantum computation. A widely separated pair of Majorana bound states forms a nonlocal fermionic state which is immune to local sources of decoherence, thus providing a platform for fault-tolerant quantum memory.
Moreover, since Majorana states realize a representation of the non-Abelian braid group, topological quantum information processing can, in principle, be effected by braiding operations~\cite{Nayak08}.
A realization of such states where they can be readily moved around and manipulated is therefore highly desirable.

There are several suggestions for physical systems that support
Majorana states, for particular ways to manipulate them, and for
measuring their properties. These include fractional quantum Hall
states at filling factor $\nu=5/2$~\cite{Moore91}, $p$-wave
superconductors~\cite{Rice95}, surfaces of three-dimensional
topological insulators in proximity to a superconductor~\cite{Fu08},
and helical edge modes of two-dimensional topological insulators in
proximity to a ferromagnet and a
superconductor~\cite{Nilsson08}. Recently, it was suggested that a
semiconducting thin film sandwiched between an $s$-wave superconductor
and a magnetic insulator~\cite{Sau10} will host Majorana states
associated with superconducting vortices. All these proposals,
however, would be experimentally extremely challenging.

Realizing and manipulating Majorana fermions in wires may be
decisively simpler. Here we show that quantum wires with strong
spin-orbit coupling, such as InAs or InSb wires, and banded carbon nanotubes
form a helical liquid, similar to the edges of a topological
insulator. Consequently, these wires support
Majorana states when they are in proximity to $s$-wave
superconductors, and a magnetic field. Most importantly, we explain
how they can be produced and manipulated by variations of a chemical
potential, which could be simply produced by a set of micron-sized gates
capacitatively coupled to the wire. Below we outline the key physical
properties of Majorana states in quantum wires, their experimental
signatures, and how Majorana-supporting wires could be extended into
networks of Majorana fermions, enabling quantum information processing.

Our analysis begins with writing the Hamiltonian for a spin-orbit coupled quantum
wire. Without loss of generality, let us choose the wire to lie along
the $y$ direction, the spin-orbit interaction, $u$, to be along the $z$
direction, and a magnetic field $B$ to be along the $x$ direction.
In addition, the wire is in contact with a superconductor,
with the proximity strength being $\Delta$ (assumed to be real). The Hamiltonian is given by~\cite{Fu08}
\begin{eqnarray}
\label{eq:H0}
H&=&\int \Psi^\dagger(y) {\cal H} \Psi(y) dy\; ; \;\;  \Psi^\dagger=\left(\psi^\dagger_\uparrow,\psi^\dagger_\downarrow,\psi_\downarrow,-\psi_\uparrow \right) \\
{\cal H}&=&\left[p^2/{2m}-\mu(y)\right]\tau_z+ u(y) p\; \sigma_z \tau_z+B(y)\sigma_x+\Delta(y)\tau_x. \nonumber
\end{eqnarray}
Here, $\psi_{\uparrow,(\downarrow)}(y)$ is the annihilation operator
of electrons with up (down) spin at position $y$.  The Pauli matrices
$\sigma$ and $\tau$ operate in spin and particle-hole space,
respectively. $\mu$ is the chemical potential.

In the absence of the Zeeman field and the superconducting proximity,
the eigenstates of the Hamiltonian (\ref{eq:H0}) will have an energy-momentum dispersion consisting of two
shifted parabolas crossing at momentum $p=0$. The Zeeman field $B$ removes the
level crossing and  opens a gap at $p=0$. We note that such a gap may also occur due to strong
electron-electron interactions \cite{Xu06,Wu06}, and therefore $B$
should be generally contrued as either a magnetic field perpendicular
to the spin orbit coupling, or an interaction induced gap.  The pairing
$\Delta$ will play two crucial roles:  Opening a gap at the outer
wings of the dispersion, where the Zeeman field is unimportant, and
modifying the gap forming near $p=0$. The former role eliminates the
possibility of high-momentum gapless excitations, thus leaving only the
chiral states near $p=0$ as low energy excitation. These states
resemble the edge of a topological insulator \cite{Wu06,Fu08}. The latter
role allows us to tune the topological phase transitions essential for the
production of Majorana fermions.  Note that another way for
gapping out the large momentum excitations is by coupling our system
to an antiferromagnet with periodicity comparable to $2 k_F$ of the
wire. Interactions may also open a pairing $\pm k_F$ gap for
chemical potentials away from the Zeeman gap \cite{Sun07}.

\begin{figure}[b]
\begin{center}
\includegraphics*[width=.2\textwidth]{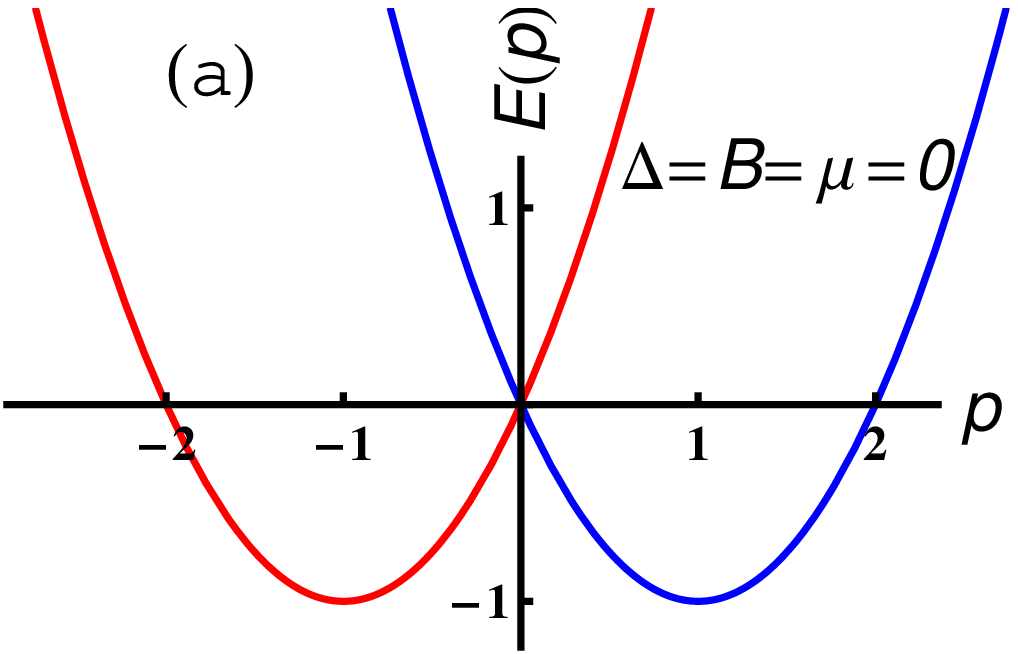}
\includegraphics*[width=.2\textwidth]{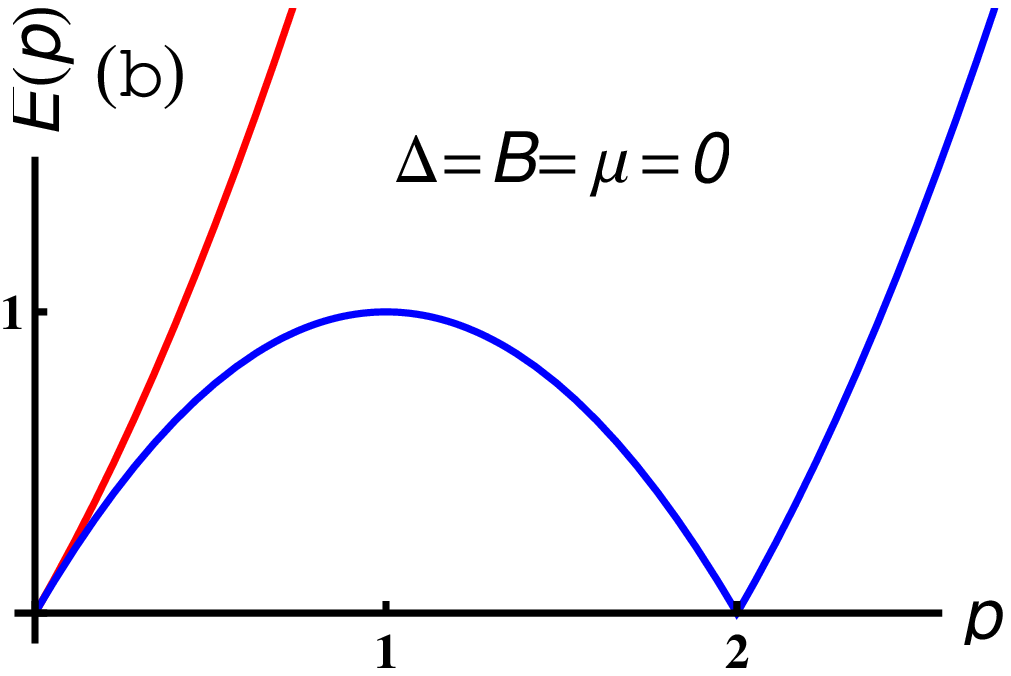}
\includegraphics*[width=.2\textwidth]{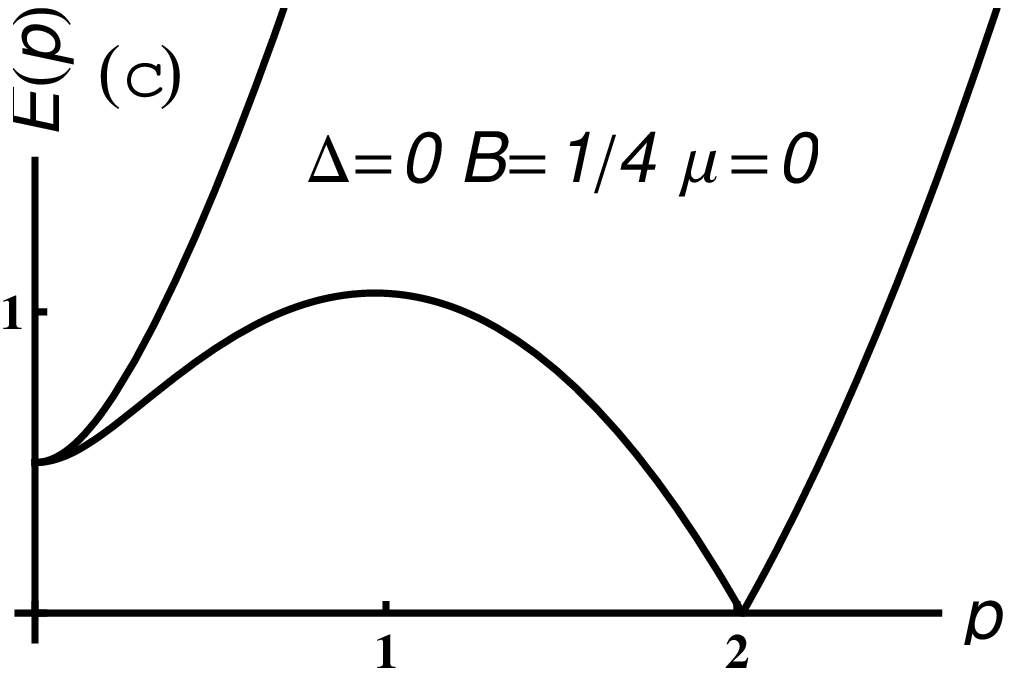}
\includegraphics*[width=.2\textwidth]{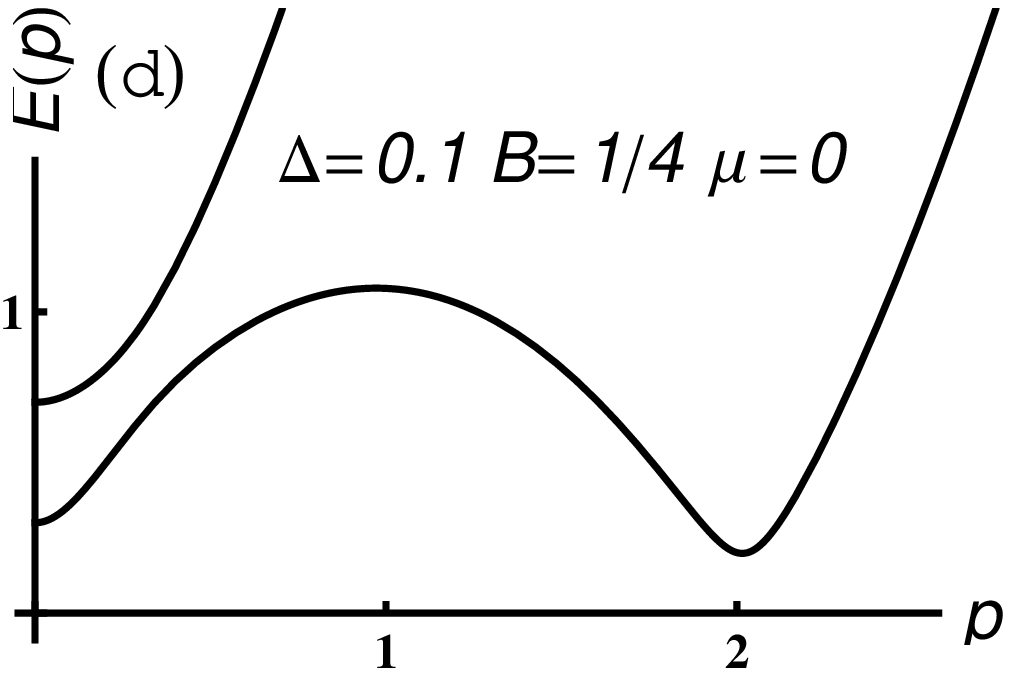}
\includegraphics*[width=.2\textwidth]{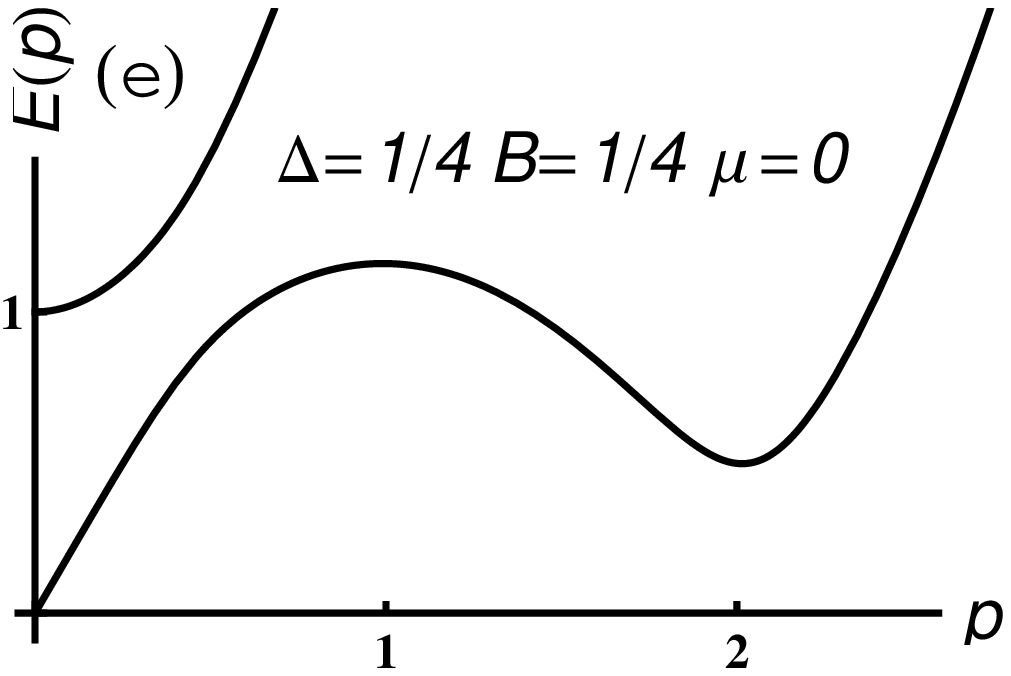}
\includegraphics*[width=.2\textwidth]{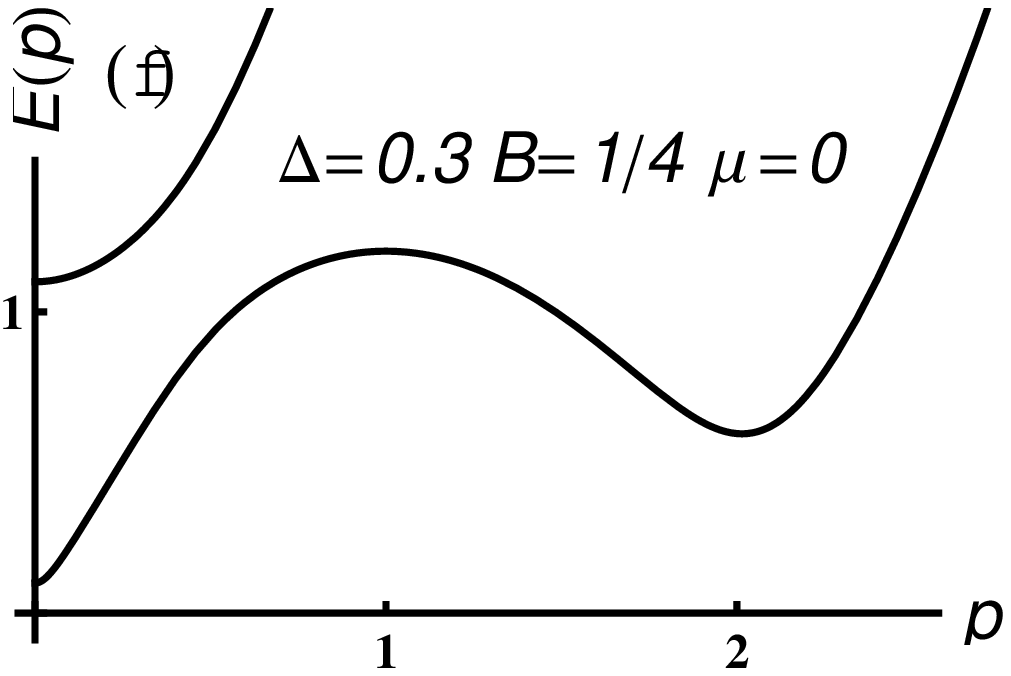}
\includegraphics*[width=.2\textwidth]{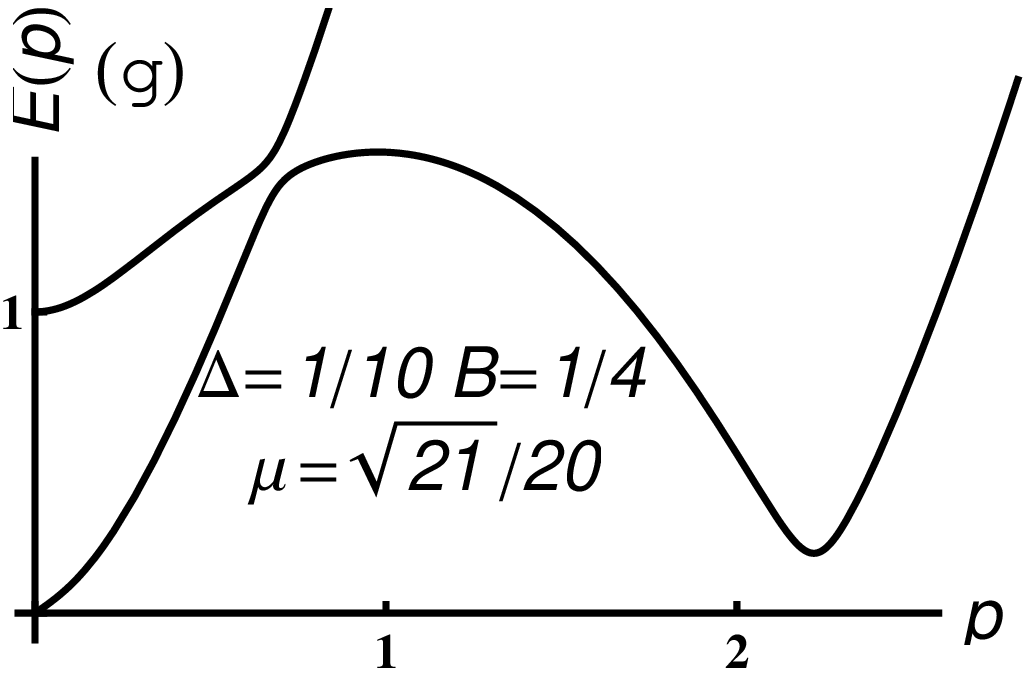}
\includegraphics*[width=.2\textwidth]{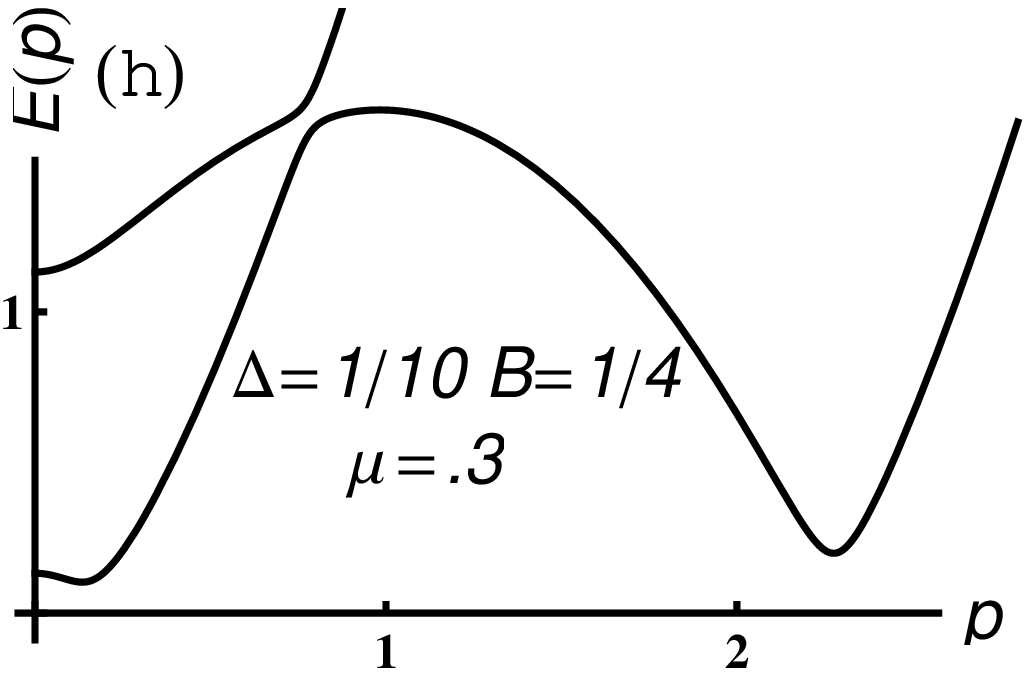}
\end{center}
\caption{\label{fg:spectrum} (a) Single-particle spectrum for $\mu=B=\Delta=0$. (The two colors denote the different spin components). The energy (momentum) scale is set by $m u^2/2$ (by $m u$), with $u$ the spin-orbit coupling strength. (b) Excitation spectrum of adding or removing an electron for $\mu=B=\Delta=0$. (c) Excitation spectrum for $B=1/4$, $\Delta=\mu=0$ where a spin gap opens near $p=0$.  (d) $B=2.5, \Delta = 1/2$, $\mu=0$ with a superconducting gap in the wings and a spin gap near the origin. This situation is analogous to a $p$-wave superconductor. We refer to this phase as the "spin gap phase" (e) $B=1/4=\Delta=1/4$, $\mu=0$. The gap near $p=0$ closes, the gap at finite $p$ persists. At this critical point a quantum phase transition occurs. (f) $B=1/4, \Delta=0.3$, $\mu=0$. All gaps in the excitation spectrum are controlled by $\Delta$. (g) $B=1/4, \Delta = .1,\mu=\sqrt{B^2-\Delta^2}=\sqrt{21}/20$. The gap at $p=0$ closes due to the shift in chemical potentia
 l. (h)
A superconducting gap opens up in the entire spectrum due to the shift of the chemical potential above its critical value $\Delta=1/10, B=1/4,\mu=0.3$.}
\end{figure}

The emerging spectrum for constant $\mu$, $u$, $\Delta$, and $B$, is
conveniently obtained by squaring the Hamiltonian twice. This straightforwardly yields the expression:
\begin{equation}
\label{eq:Eigen}
 E_\pm^2=B^2+\Delta^2+\xi_p^2+(u p)^2 \pm 2 \sqrt{B^2 \Delta^2+B^2 \xi_p^2+(u p)^2 \xi_p^2}
\end{equation}
where $\xi_p=p^2/2m-\mu$. Fig.~\ref{fg:spectrum} displays the spectrum
for several values of $B$, $\Delta$, and $\mu$.
As these parameters
vary (while $B$ and $\Delta$ remain nonzero), a gap closing and reopening indicates a topological phase
transition. Generically, we expect gaps appearing near $p=0$ and near
the Fermi momenta corresponding to $\xi_p \pm u p =0 $.  We will denote
these gaps as $E_0$ and $E_1$, respectively.

As hinted above, it is the zero-momentum gap, $E_0$, which is crucial for our understanding of
the emerging Majorana states. Examining $E_-$ at $p=0$ we notice that
\begin{equation}
\label{eq:E0}
E_0=E(p=0)= |B-\sqrt{\Delta^2+\mu^2}|.
\end{equation}
For $B^2> \Delta^2+\mu^2$,  $E_0$ is a spin gap due to the Zeeman
field (or strong interaction), while for $B^2< \Delta^2+\mu^2$ it is a
superconducting gap, thus when  $B^2= \Delta^2+\mu^2$ a quantum phase
transition occurs. At the same time the gap $E_1$ near $p^2=2 \mu m$
is always a superconducting gap, as we require $\Delta$ to always
remain finite.

\begin{figure}[h]
\begin{center}
\includegraphics*[width=.4\textwidth]{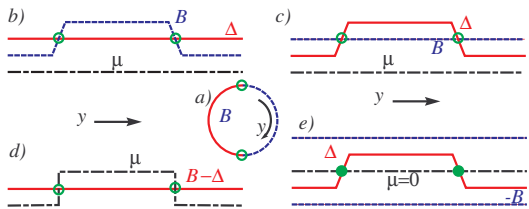}
\end{center}
\vspace{-.5cm}
\caption{\label{fg:maj} (a) Wire in a ring geometry. Both halves have constant parameters and are joined by short junctions with a linearly varying parameter. Majorana states (marked by circles) are formed at the junctions. (b) Majorana state in the sector $p=0$ when $B$ varies. The gap in the finite-$p$ sector remains finite in the entire wire. (c) Majorana state in the sector $p=0$ when $\Delta$ varies. (d) Majorana state in the sector $p=0$ when $\mu$ varies. (e) ``$p$-wave" Majorana state when $\Delta$ changes sign. The sector $p=0$ remains gapped in the entire wire. Each crossing with $\Delta=0$ hosts two Majorana states.}
\end{figure}

The phase transition evident in $E_0$ allows the formation of {\it
  Majorana states}. Indeed, the dependence of $E_0$ on $B$, $\Delta$,
and $\mu$ enables us to construct zero-energy Majorana states in
various ways. As in edge states of 2D topological insulators
\cite{Fu08}, a Majorana bound state will form when $B$ changes
in space and crosses $\Delta$, e.g. at $y=0$
(cf.\ Fig~\ref{fg:maj}b), or when $\Delta$
varies in space and crosses $B$ (cf. Fig~\ref{fg:maj}d).

Here we emphasize, however, a third possibility: varying the chemical
potential, $\mu$. Let us assume that $B>\Delta$ so that for $\mu=0$ we have a
spin gap $E_0$. But when $\mu >\sqrt{B^2-\Delta^2}$, the gap $E_0$,
Eq.~(\ref{eq:E0}), is clearly superconducting. Thus, we can form a Majorana state by
tuning $\mu$ between these two values (cf.\ Fig~\ref{fg:maj}c). We
note that changes in $\mu$ do not significantly influence the gap
$E_1$, so that the electronic states near $\pm k_F$ do not play a
role.

The one-dimensional geometry allows for a simple demonstration of how
to form Majorana states where their wave functions can be obtained
essentially exactly. Let us consider these examples in a long ring with one conducting
channel, in proximity to a superconductor and a Zeeman field, as
illustrated in Fig.\ \ref{fg:maj}a. Since the relevant momenta are
near $p=0$, in the treatment below we use the Hamiltonian linearized in
that region:
\begin{equation}
\label{eq:B}
{\cal H}= u p \; \sigma_z \tau_z-\mu(y)\tau_z + B(y) \sigma_x + \Delta(y)\tau_x
\end{equation}

{\bf Spatially varying $B$.} Assume $\Delta>0$ is constant, $\mu=0$, and that $B>\Delta$ for $y>0$ and
$B<\Delta$ for $y<0$  (Fig.\ \ref{fg:maj}b; note that the periodic
boundary conditions require another point where $B=\Delta$).
Near the crossing point $y=0$, we write $B(y) = \Delta + b y$.
Due to particle-hole symmetry, it is useful to square the Hamiltonian
Eq.~(\ref{eq:B}) to diagonalize it. In addition to the square of each term and the mixed $B \Delta$ term, we obtain a term $\{u p \sigma_z \tau_z, B \sigma_x \}= i \sigma_y \tau_z u [p,B]= \sigma_y \tau_z u b $ which arises because $B$ depends on space and does not anticommute with the spin-orbit coupling. Collecting all terms, we have
\begin{equation}
\label{eq:H2B}
  {\cal H}^2_b = (u p )^2 +B(y)^2+\Delta^2+ u b \sigma_y \tau_z +2 \Delta B(y) \sigma_x \tau_x
\end{equation}
Rotating ${\cal H}_b^2$ by $U^\dagger_b=1/2\left(\tau_z-i \tau_x - i \sigma_x \tau_z+ \sigma_x \tau_x \right)$, we find that $U_b\cdot{\cal H}_b^2\cdot U_b^\dagger$ is diagonal with components $(u p)^2+\left(\Delta\pm B\right)^2 \pm u b$.
The interesting modes are those with a minus sign in the brackets,
$\Delta-B$. They correspond to a simple harmonic oscillator
Hamiltonian with ground-state wave function $\varphi(y)=( b/(u
\pi)^{1/4})  e^{- b y^2/ (2 u)}$  and energies
$E^2_n= 2 u b (n +1/2) \pm u b,\; n=0,1,2,\ldots$. For $b>0$, the minus sign yields a zero-energy state with Bogoliubov operator
\begin{eqnarray}
\label{eq:sol}
 \gamma^\dagger_b&=& \gamma_b= \frac{1}{\sqrt{2}} \left(\eta_1 - \eta_2\right)=\frac{1}{2}\left( \psi_\uparrow - i \psi_{\downarrow}+ i \psi^\dagger_\downarrow+\psi^\dagger_\uparrow \right) \nonumber \\
\eta_1&=&1/\sqrt{2}\left({\psi^\dagger_\uparrow+\psi_\uparrow}\right); \;
\eta_2=1/(\sqrt{2} i ) \left({\psi^\dagger_\downarrow-\psi_\downarrow}\right).
\end{eqnarray}
The Majorana state at the second crossing point along the ring follows by $b \rightarrow -b$. Thus, this zero-energy state is $E^+_0=0$ with Majorana operator $-i/\sqrt{2} (\eta_1+\eta_2)$.

{\bf Spatially varying $\Delta$.} For the case where  $\Delta$ depends on $y$,
we assume $\Delta(y)= B + d y$, $\mu=0$, and a constant $B$ (Fig.\ref{fg:maj}c). The Hamiltonian here is similar to that in the y-dependent $B$ case, if we exchange $\tau$ and $\sigma$ in
Eqs. (\ref{eq:B}) and (\ref{eq:H2B}). Therefore, the Majorana states emerge
in this case in exactly the same way as above, except with the
diagonalizing matrices being $U^\dagger_d= U^\dagger_b (\tau
\leftrightarrow \sigma)$, and with $b$ and $\Delta$ exchanged with $d$
and $B$ respectively in the resulting wave function. This yields (for positive $d$) $\gamma_d=\gamma_d^\dagger=\left(\eta_1-\eta_2\right)/\sqrt{2}$.

{\bf Spatially varying $\mu$}. If $B>\Delta$ in the entire wire, then
at the interface between spin-gap regions with $\mu^2 < B^2-\Delta^2$
and pairing gap regions with $|\mu|^2> B^2-\Delta^2$, a Majorana state
will also form (Fig.\ref{fg:maj}d). In this case, we assume that
$\mu$ jumps abruptly at $y=0$ between $\mu_{\ell}$ for $y<0$, and
$\mu_r$ at $y>0$. The condition for the Majorana state to form is:
\begin{equation}
\mu_{\ell}^2 < B^2-\Delta^2,\,\,\, \mu_r^2 > B^2-\Delta^2
\end{equation}
We match the wave function
at $y=0$, using the ansatz $\psi_r \propto e^{-k_r y} $ for $y>0$ and
$\psi_l \propto e^{k_l y} $ for $y<0$. The Hamiltonian becomes:
\be
{\cal H} = -
(\Theta(y)k_r-\Theta(-y)k_l)i u \tau_z \sigma_z - \mu \tau_z +
B\sigma_x +\Delta \tau_x=0
\ee
where $u k_{r,(l)\pm}= \Delta \pm
\sqrt{B^2-\mu_{(l)r}^2}$ and the eigenvectors
\be
\psi_\pm^r=e^{-(+)k_{r(l) \pm} y}\left(1,e^{\pm i\theta_{r(l)}},i,-i
e^{\pm i \theta_{r(l)}} \right)^T/2
\ee
with $e^{i\theta_{r(l)}}=
\mu_{r(l)}/B+i \sqrt{1-\mu_{r(l)}^2/B^2}$. It is straightforward to verify
that $\psi_{r(l)}\cdot \Psi =(\psi_{r(l)}\cdot \Psi)^\dagger$ are
Majorana operators, with $\psi$ a simple c-number. Thus, we find that
the wave function $\psi(y)$ of the Majorana state is
\begin{equation}
\label{eq:mu}
\hspace{-.065cm}
\left\{\begin{array}{lc}
2i\sin\theta_r\cdot\psi^{(0)\ell}_{-} & y<0 \\
(e^{-i\theta_l}-e^{-i\theta_r})\psi^{(0)r}_{+}+
(e^{i\theta_r}-e^{-i\theta_l})\psi^{(0)r}_{-} &
y>0
\end{array}\right.,
\end{equation}
which exhausts all possibilities for isolated majorana states.

Indeed we must note that when $E_0$ is a spin gap, the gap $E_1$ is
due to pairing between spin-up electrons for positive $p$ and
spin-down electrons for negative $p$, reminiscent of a one-dimensional
$p$-wave superconductor \cite{Notepwave10}. Recalling that vortices of a $p$-wave
superconductor support a zero-energy bound state~\cite{Moore91,Read00,Sau10}, we
expect the formation of Majorana states when $\Delta$ changes
sign (Fig. \ref{fg:maj}e). Due to the broken azimuthal symmetry, however, two inseparable Majorana
states form where $\Delta$ vanishes.

Next we discuss {\it experimental realizations}. The main requirement
for our proposal to be feasible is a sufficiently strong spin-orbit
interaction. Spin-orbit coupling in wires adiabatically connected to
reservoirs was considered long ago, both without electron-electron
interactions~\cite{Moroz99} and with interactions~\cite{Moroz00} in
the framework of Luttinger-liquid theory. Recently, this problem
attracted renewed theoretical \cite{Hattori06} and experimental
\cite{Wan09} interest, both with and without external magnetic field.

Several candidate systems for quantum wires with spin-orbit
interaction exist. In carbon
nanotubes, spin-orbit coupling arises due to curvature
effects~\cite{Kuemmeth08}. Here it is preferable to have a strong
spin-orbit coupling along
the direction of propagation, requiring that the tube is
bent along its axis. Alternatively, one could introduce a strong
electric field perpendicular to the axis. Perhaps a more promising candidate is a wire
of InAs in the wurtzite structure which is known to have strong
spin-orbit coupling~\cite{Fasth07}. The velocity $u$ in the
Hamiltonian Eq. (\ref{eq:H0}) is related to the
experimentally measured length scale $\lambda_{\rm S0} = 100
nm = m u$ and $\Delta_{\rm SO} = 250 \mu V= m u^2/2$ via $u \sim \hbar
2 \Delta_{\rm SO} \lambda_{\rm SO} \approx 7.6 \times 10^6$cm/sec and
$m= \hbar^2/\lambda_{\rm SO}^2 2 \Delta= 0.015 m_e$, with $m_e$ the
free electron mass. Similar numbers (with $\Delta=280\mu V$) describe newly fabricated InSb wires, except with a large g-factor of $\sim 50$, compared to $g\sim 8$ in InAs, requiring only a small, relatively innocuous to the SC, magnetic field\cite{Nilsson09}.

The wire-Majorana states we envision, can be formed by spatial variations of the
Zeeman field, the proximity-induced superconductivity, or, most
importantly, the chemical potential, and will form near points where
$B^2-(\mu^2+\Delta^2)=0$. A varying chemical
potential, as in Fig~\ref{fg:maj}d, for instance, can be achieved by
gate electrodes capacitatively coupled to the wire. Tunneling experiments should provide the most direct signatures
of the Majorana states \cite{Kraus09}.

\begin{figure}[t]
\begin{center}
\includegraphics*[width=.45\textwidth]{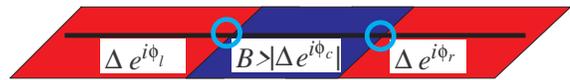}
\end{center}
\vspace{-1cm}
\caption{\label{fg:var} $\mu$, $B$ or $\Delta$ are tuned so that in the middle superconductor we have a spin gap while the other superconductors have a superconduting gap. In this configuration, it is possible to manipulate the two junctions separately by changing the superconducting phase difference between the neighboring regions.}
\end{figure}

Additional experimental signatures can be probed by controlling the
phase of the pairing $\Delta$ in addition to the chemical potential.
In particular, the configuration of Fig.~\ref{fg:var} allows controlling the pairing phase on
the left, center, and right sections independently; we denote these
phases by $\phi_{\ell},\,\phi_c,\,\phi_r$. The total Joephson
current flowing between the three superconducting segments is rather
intricate, and will be discussed in a separate publication. Since the Majoranas are localized when the distance between them,  $L$, is infinite the Jospehson current due to the Majoranas is zero.
A straightforward first-order perturbation analysis for finite $L$ yields the energy
splittings between the two Majorana states on the domain walls (c.f. Ref. \cite{Shivamoggi}). We
find the Josephson energy associated with the Majorana fermions to be:
\be
E=E_{\ell r}\cos(\frac{\phi_{\ell}-\phi_{r}}{2})+E_{c}\cos(\frac{\phi_{\ell}+\phi_{r}}{2}-\phi_c).
\label{JJmaj}
\ee
here we assume that $\mu_c=0$ in the center region, and
$\mu_{\ell}=\mu_r=\mu$ on the sides. Also, $E_{\ell r}\sim E_{c}\approx \sqrt{\frac{2 \Delta B(\Delta^2-B^2)(\Delta^2+\mu^2-B^2)}{\Delta^2(\Delta^2+\mu^2-B^2)-B\mu^2(B+\Delta)}}e^{-(B-\Delta)L/u}$. In the similar
setup of the edges of a topological insulator \cite{Fu09a,Akhmerov09} the $E_c$, which is a result of the tunneling
  from the left and the right sections to the middle section is absent since the
center region between the Majoranas is not in proximity to a
superconductor. We notice two prominent effects. First, by letting
$\phi_c=\phi_r$, for instance, we see that the Josephson current from
the left superconductor is $4\pi$ periodic. More interestingly, when
we try to draw current from the center region, the current,
proportional to $E_c\sin((\phi_{\ell}+\phi_r-2\phi_c)/2)$ is drawn
equally from the left and right superconductors. Therefore this
geometry can serve as a Josephson transistor, since a change of $\phi_R$ determines part of the current between the left and the middle section. Related effects were
discussed in Ref.~\cite{NagaosaSFS}.

Next, let us estimate the Josephson tunneling strengths in the setup
of Fig. \ref{fg:var}, with three independent superconducting substrates. If the substrates are properly insulated from each other, only Josephson currents will be carried through the
quantum wire via proximity; these will be given by Eq. (\ref{JJmaj}), in addition to a contribution
proportional to $\Delta$ which is $2\pi$ periodic in $\phi_r-\phi_c$
and $\phi_{\ell}-\phi_c$. For the energy scales associated with the
$InAs$ wires, we expect the critical current for the $2\pi$ periodic
portion to be of order $40nA$, consistent with $B,\Delta\sim 1K$. The
$4\pi$-periodic critical Majorana-Josephson currents,
$2eE_c/\hbar$ and $2e E_{\ell r}/\hbar$, are a significant fraction of
this number. For instance, for $InAs$ parameters with $\mu_c=0$ and
$\mu_r=\mu_{\ell}=0.9B$ and $\Delta=0.8 B$, with $B\sim 1K$, we obtain
$E_c\approx 0.22K e^{-L (B-\Delta)/u}$, corresponding to a maximum current $e
E_c/\hbar\approx 4 nA\cdot e^{-L/3\mu m}$ with $L$ the separation
between the Majoranas. The unique flux periodicity of the Majorana-Josephson currents can also be
probed with low-frequency shot noise would also reveal the
anomalous Josephson periodicity.


In this manuscript we have shown that wires with strong spin-orbit
coupling in proximity to a superconductor host an interesting
effective helical state. By tuning the superconducting gap $\Delta$,
the spin gap $B$, or the chemical potential $\mu$,  Majorana states
can be created and detected in various experimental ways.
By fabricating a set of gates over a network of such wires, we can
imagine adiabatically creating Majorana pairs, moving, and even interchanging them along the network using pulse
sequences in the gates. The non-Abelian character of the system should become
apparent in such networks. Methods to manipulate the Majorana modes
and experimental consequences for the conductance will be the subject
of a future manuscript.

While finishing this manuscript we became aware of the preprint \cite{Lutchyn10} which has some overlap with our results.

{\it Acknowledgements}.--- We would like to thank Joel Moore, Jason
Alicea, Erez Berg and Oleg Starykh for enlightening discussions. The
research was supported by ISF, DIP and BSF grants, as well as Packard and Sloan
fellowships and from the Institute for Quantum Information under NSF
grants PHY-0456720 and PHY-0803371.
\bibliographystyle{prsty}
\bibliography {F:/oreg/MyDocs/G-disk-Active/ref/library}

\begin{thebibliography}{10}
\bibitem{Notepwave10}
Similar to the case of topological insulators the analogy can be made precise
  by a unitary transformation \protect{\cite{Fu08}}.
\end{thebibliography}

\end{document}